\documentclass[a4paper,11pt]{amsart}
\usepackage{graphicx}
\usepackage{amssymb}
\begin{document}

\hyphenation{gra-vi-ta-tio-nal re-la-ti-vi-ty Gaus-sian
re-fe-ren-ce re-la-ti-ve gra-vi-ta-tion Schwarz-schild
ac-cor-dingly gra-vi-ta-tio-nal-ly re-la-ti-vi-stic pro-du-cing
de-ri-va-ti-ve ge-ne-ral ex-pli-citly des-cri-bed ma-the-ma-ti-cal
de-si-gnan-do-si coe-ren-za pro-blem gra-vi-ta-ting geo-de-sic
per-ga-mon cos-mo-lo-gi-cal gra-vity cor-res-pon-ding
de-fi-ni-tion phy-si-ka-li-schen ma-the-ma-ti-sches ge-ra-de
Sze-keres con-si-de-red tra-vel-ling ma-ni-fold re-fe-ren-ces
geo-me-tri-cal in-su-pe-rable sup-po-sedly at-tri-bu-table
Bild-raum in-fi-ni-tely counter-ba-lan-ces iso-tro-pi-cally}

\title[On SgrA* -- Theory and Observations]
{{\bf On Sagittarius A*\\Theory and Observations}}

\author[Angelo Loinger]{Angelo Loinger}
\address{A.L. -- Dipartimento di Fisica, Universit\`a di Milano, Via
Celoria, 16 - 20133 Milano (Italy)}
\author[Tiziana Marsico]{Tiziana Marsico}
\address{T.M. -- Liceo Classico ``G. Berchet'', Via della Commenda, 26 - 20122 Milano (Italy)}
\email{angelo.loinger@mi.infn.it} \email{martiz64@libero.it}

\vskip0.50cm

\begin{abstract}
Massive and supermassive ``dust'' spheres (with a zero internal
pressure) collapse to ``full globes'' of finite volumes, whose
surfaces have the properties of the event horizon around a
mass-point. This fact explains the observational data concerning
Sagittarius A* (SgrA*). By virtue of Hilbert's repulsive effect,
both the event horizon of a mass-point and the event horizon of a
``full globe'' cannot ``swallow'' anything.
\end{abstract}

\maketitle

\vskip1.20cm \noindent \small \textbf{Summary} -- \textbf{1},
\textbf{2}. Some observational data about SgrA*. -- \textbf{3}.
The geodesics of Schwarzschild's manifold created by a point-mass
and the gravitational repulsion. -- \textbf{3bis}. Inadequacy of
the proper time as evolution parameter of the geodesic motions
in a Schwarzschild's manifold. -- \textbf{4}. Kerr's manifold and
gravitational repulsion. -- \textbf{5}, \textbf{5bis}. A
`dust'' sphere collapses to a `full globe'' of a finite volume;
\emph{et cetera}. -- \textbf{6}. Explanation of the observational
data about SgrA*. -- \emph{Appendix A}: On celestial objects
endowed with a magnetic moment. -- \emph{Appendix B}: Attraction
and repulsion in Hilbert's and Droste's treatments. --

\normalsize

\vskip1.20cm \noindent \textbf{1.} -- In the paper ``The event
horizon of Sagittarius A*'' by Broderick \emph{et al.} \cite{1},
the authors affirm that recent millimeter and infrared
observations of the supermassive centre SgrA* of the Milky Way
require the existence of the ideal surface of an event horizon. In
the present paper we show that the observational data of \cite{1}
are perfectly consistent with the existence of a \emph{real},
peculiar surface, which has all the properties of the above event
horizon. Remark that the horizons have \emph{no} ``swallowing''
property -- a fact which is commonly ignored.

\vskip1.20cm \noindent \textbf{2.} -- According to Broderick
\emph{et al.} \cite{1}, by virtue of its relative proximity SgrA*
is the best candidate to the possession of the ideal surface of an
event horizon.

\par Observations of massive stars in its vicinity have given the
following values for its mass and its distance: mass
$M=(4.5\pm0.4)\times10^{6}$ solar masses; distance $D=(8.4\pm0.4)$
kpc; SgrA* is confined within 40AU. The radiative emission
(luminosity of $10^{36}$ erg s$^{-1}$) is strongly non-thermal, and
is distributed from the radio to the $\gamma$-rays.

\par The substance of the arguments of Broderick \emph{et al.}
\cite{1} is admirably summarized in the first paragraph of the
final sect.\textbf{4}, which we report literally: ``Recent
infrared and mm-VLBI observations imply that if the matter
accreting onto Sgr A* comes to rest in a region visible to distant
observers, the luminosity associated with the surface emission
from this region satisfies $L_{\rm surf}/L_{\rm acc} \lesssim
0.003$. Equivalently, these observations require that $99.6\%$ of
the gravitational binding energy liberated during infall is
radiated in some form prior to finally settling.  These numbers
are inconsistent by orders of magnitude with our present
understanding of the radiative properties of Sgr A*'s accretion
flow specifically and relativistic accretion flows generally.
Therefore, it is all but certain that no such surface can be
present, i.e., {\em an event horizon must exist}.''

\par Now, if we take into account the decisive role of the
Hilbertian gravitational \emph{repulsion} \cite{2}, \cite{3}, which
is neglected by our authors \cite{1}, the picture changes
drastically, and the above inequality $L_{\rm surf}/L_{\rm acc}
\lesssim 0.003$ becomes quite comprehensible, as we shall see in
the sequel.

\vskip1.20cm \noindent \textbf{3.} -- For the computation of the
geodesics of the Schwarzschild manifold created by a material
point, Hilbert \cite{2} starts from the standard
(Hilbert-Droste-Weyl) form of the interval $\textrm{d}s$:

\begin{equation} \label{eq:one}
\textrm{d}s^{2} = \frac{r}{r-\alpha} \, \textrm{d}r^{2} + r^{2}
\textrm{d}\vartheta^{2} + r^{2} \sin^{2}\vartheta \,
\textrm{d}\varphi^{2}- \frac{r-\alpha}{r} \, \textrm{d}t^{2} \quad
; \quad (c=G=1) \quad,
\end{equation}

where $\alpha\equiv 2m$, and $m$ is the mass of the gravitating
point -- if $M$ is the mass in CGS units, we have $M=c^{2}m/G$.
(The original Schwarzschild's form of $\textrm{d}s^{2}$ can be
obtained from eq. (\ref{eq:one}) with the substitution
$r\rightarrow (r^{3}+\alpha^{3})^{1/3}$.)

\par It is easy to see that there are only \emph{plane}
trajectories, and therefore it suffices to consider only one value
for $\vartheta$, \emph{e.g.} $\pi /2$. Eq. (\ref{eq:one}) has as an
evident consequence the following first integrals of the geodesic
motions, where $A, B, C$ are constants with respect to the affine
parameter $p$:

\begin{equation} \label{eq:two}
\frac{r}{r-\alpha} \, \left(\frac{\textrm{d}r}{\textrm{d}p}
\right)^{2} +r^{2}  \left( \frac{\textrm{d}\varphi}{\textrm{d}p}
\right)^{2} - \frac{r-\alpha}{r} \, \left(
\frac{\textrm{d}t}{\textrm{d}p} \right)^{2} = A \quad;
\end{equation}

\begin{equation} \label{eq:three}
r^{2} \, \frac{\textrm{d}\varphi}{\textrm{d}p} = B \quad;
\end{equation}

\begin{equation} \label{eq:four}
\frac{r-\alpha}{r} \, \frac{\textrm{d}t}{\textrm{d}p}= C \quad .
\end{equation}

Clearly, $A$ is negative for the test-particles and zero for the
light-rays. With a suitable choice of $p$, we can put $C=1$. Then,
by eliminating $t$ and $p$ from eqs.
(\ref{eq:two})--(\ref{eq:three})--(\ref{eq:four}), we obtain the
\emph{general formula} of \emph{all} the geodesic lines:

\begin{equation} \label{eq:five}
\left( \frac{\textrm{d}\varrho}{\textrm{d}\varphi} \right)^{2} =
\frac{1+A}{B^{2}} - \frac{A\alpha}{B^{2}} \, \varrho - \varrho^{2}
+ \alpha\varrho^{3} \, \Big[=
\left(\frac{\textrm{d}r}{\textrm{d}p}\right)^{2} \,
\frac{1}{B^{2}}\Big] \quad ,
\end{equation}

where $\varrho\equiv 1/r$. Remark that the coordinate $r$ in eq.
(\ref{eq:five}) must satisfy the condition $r>\alpha$, because the
progenitor eqs. (\ref{eq:two}) and (\ref{eq:four}) do not hold for
$r\leq \alpha$. Remark that by substituting $\textrm{d}p=
[(r-\alpha)/r]\,\textrm{d}t$ in eqs.
(\ref{eq:two})--(\ref{eq:three}) one obtains two first integrals
with respect to the evolution parameter $t$.

\par Circular and radial orbits are evidently possible; however,
for the circular motions it is necessary to use also the
Lagrangean equation of motion for $r$:

\begin{equation} \label{eq:six}
\frac{\textrm{d}}{\textrm{d}p} \left(\frac{2r}{r-\alpha} \,
\frac{\textrm{d}r}{\textrm{d}p} \right) +
\frac{\alpha}{(r-\alpha)^{2}} \,
\left(\frac{\textrm{d}r}{\textrm{d}p}\right)^{2} - 2r
\left(\frac{\textrm{d}\varphi}{\textrm{d}p}\right)^{2} +
\frac{\alpha}{r^{2}} \,
\left(\frac{\textrm{d}t}{\textrm{d}p}\right)^{2} = 0 \quad ,
\end{equation}

since, when $\textrm{d}r / \textrm{d}p=0$, this equation is
\emph{not} an analytical consequence of eqs.
(\ref{eq:two})--(\ref{eq:three})--(\ref{eq:four}).

\par One finds that the velocity $v=r\, \textrm{d}\varphi /
\textrm{d}t$ on a \emph{circular} orbit is given by

\begin{equation} \label{eq:seven}
v^{2}= \left(\frac{r\,\textrm{d}\varphi}{\textrm{d}t}\right)^{2} =
\frac{\alpha}{2r} \quad ;
\end{equation}

for the test-particles we have that

\begin{equation}\label{eq:eigth}
v < \frac{1}{\sqrt{3}} \quad ,
\end{equation}

and

\begin{equation} \label{eq:nine}
r >  \frac{3\alpha}{2} \quad ,
\end{equation}

a clear example of the existence of a gravitational
\emph{\textbf{repulsion}}; for the light-rays there is a unique
circular trajectory, for which

\begin{equation} \label{eq:ten}
 v = \frac{1}{\sqrt{3}} \quad ,
\end{equation}

\begin{equation} \label{eq:eleven}
r =  \frac{3\alpha}{2} \quad ,
\end{equation}

and we see that also the light ``feels'' the gravitational
\textbf{\emph{repulsion}}. Of course, the inequalities
(\ref{eq:nine}) and (\ref{eq:eleven}) can be proved with the use
of \emph{both} the evolution parameters, $t$ and $p$. (And also
with the proper time $s$ on the circular geodetics for relation
(\ref{eq:nine})).

\par For the \emph{radial} trajectories of the geodesic motions
there is gravitational attraction where

\begin{equation} \label{eq:twelve}
\left| \frac{\textrm{d}r}{\textrm{d}t} \right| <
\frac{1}{\sqrt{3}} \, \frac{r-\alpha}{r}\quad ,
\end{equation}

and gravitational \textbf{\emph{repulsion}} where

\begin{equation} \label{eq:thirteen}
 \left| \frac{\textrm{d}r}{\textrm{d}t} \right| >
\frac{1}{\sqrt{3}} \, \frac{r-\alpha}{r} \quad .
\end{equation}

If $r^{*}$ is the value of $r$ for which $\textrm{d}^{2}r /
\textrm{d}t^{2}=0$ -- attraction and repulsion counterbalance each
other --, the velocity $\textrm{d}r / \textrm{d}t$ has its maximal
value at $r=r^{*}$:

\begin{equation} \label{eq:fourteen}
\left| \frac{\textrm{d}r}{\textrm{d}t} \right|_{\textrm{max}} =
\frac{1}{\sqrt{3}} \, \frac{r^{*}-\alpha}{r^{*}} \quad .
\end{equation}

For the light-rays we have from $\textrm{d}s^{2}=0$:

\begin{equation} \label{eq:fifteen}
\left| \frac{\textrm{d}r}{\textrm{d}t} \right| =
 \frac{r-\alpha}{r} \quad :
\end{equation}

the light is \textbf{\emph{repulsed everywhere}} by the
gravitating point-mass $m$; its velocity increases from zero at
$r=\alpha$ to $1$ at $r=\infty$.

\par Test-particles and light-rays arrive at $r=\alpha$ with $\textrm{d}r /
\textrm{d}t=0$ and $\textrm{d}^{2}r /\textrm{d}t^{2}=0$: \emph{the
spatial surface} $r=\alpha$ \emph{represents for them an
insuperable barrier}.

\par This basic fact can be also illustrated starting from eq.
(\ref{eq:five}). We consider the instance  of the light-rays
($A=0$); for the test-particles ($A<0$) the results are
qualitatively the same. The general orbit of the light-rays is
given by

\begin{equation} \label{eq:fiveprime}
\left( \frac{\textrm{d}\varrho}{\textrm{d}\varphi} \right)^{2} =
\frac{1}{B^{2}}- \varrho^{2} + \alpha\varrho^{3} \quad, \quad
(\varrho\equiv 1/r) \tag{5$'$} \quad ;
\end{equation}
\setcounter{equation}{15}

this equation has for $B=3\sqrt{3}\alpha/2$ the circle
$r=3\alpha/2$ as Poincar\'e's ``cycle'' \cite{2}. Let us
characterize a ray with the segment $B$ -- see Fig. 1 --, which
gives at infinity its distance from the vertical radial line. Fig.
1 represents intuitively some integral curves of eq.
(\ref{eq:fiveprime}) obtained with Poincar\'e's cycle theory
\cite{3}. When $B<3\sqrt{3} \alpha/2$, the light-ray arrives at
$r=\alpha$ \emph{and ends there}. When $B=3\sqrt{3} \alpha/2$, it
comes near asymptotically by spiralling to the circle
$r=3\alpha/2$. When $B>3\sqrt{3} \alpha/2$, the ray performs, in
general, several revolutions round this circle, and then goes to
infinity. Fig. 1 shows three rays of the last kind, one of them
performs a revolution. Clearly, the vertical line is characterized
by the limit $B\rightarrow 0$ -- and therefore it \emph{ends at}
$r=\alpha$.

\begin{figure}[!ht]
\begin{center}
\includegraphics[width=0.6\textwidth]{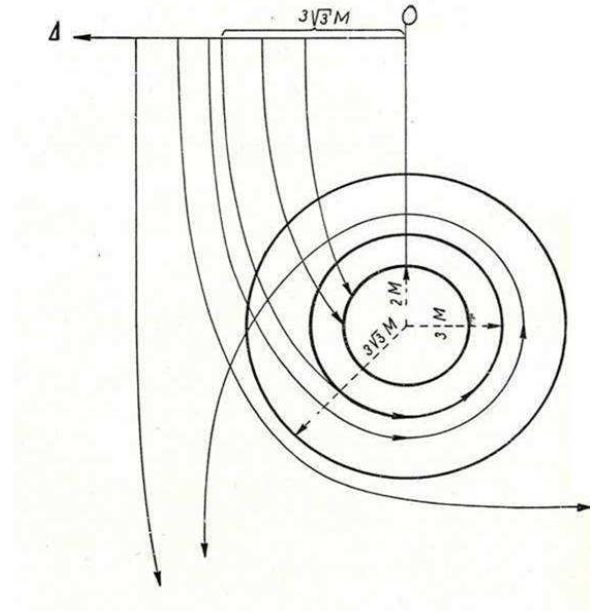}
\caption{\small See von Laue \cite{3} -- This Author writes $M$ in
lieu of our $m$, and $\Delta$ in lieu of our $B$.}
\end{center}
\end{figure}
\normalsize

\par Droste \cite{4} makes a detailed investigation of eq.
(\ref{eq:five}), whose solution is given by Weierstra\ss{}'
elliptic function. He  emphasizes that eq. (\ref{eq:one}), and all
the formulae that are a consequence of this $\textrm{d}s^{2}$, are
valid only for $r>\alpha$. In reality, the modification $(-+++)$
for $r<\alpha$ of the signature $(+++-)$ of eq. (\ref{eq:one}) --
with interchanged roles of $r$ and $t$ -- is an absurdity. Not
always does geometry coincide with physics.

\par In sect.\textbf{7} of \cite{4} we find a study of the
conditions under which there is gravitational attraction or
gravitational repulsion in the radial geodetic motions. Droste's
treatment is a little different from that of Hilbert \cite{2},
because he uses a velocity $\textrm{d}\delta / \textrm{d}t=
(\textrm{d}r / \textrm{d}t)(1-\alpha r^{-1})^{-1/2}$, which is the
time derivative of the metric distance $\delta$ of the generic
point $(r,\vartheta,\varphi)$ from $r=\alpha$. In \emph{App. B} we
give a comparison between Hilbert's and Droste's treatments.

\vskip1.20cm \noindent \textbf{3bis.} -- In sect.\textbf{3} we
have emphasized that the geodesic lines for which $B<3\sqrt{3}
\alpha/2$ arrive at $r=\alpha$ and end there. A result which
generalizes the previous conclusion that for the radial geodesics
we have $\textrm{d}r / \textrm{d}t=0=\textrm{d}^{2}r /
\textrm{d}t^{2}$ at $r=\alpha$.

\par Some theoreticians do not like the use of the ``Systemzeit''
$t$; they prefer the proper time of the test-particles (and an affine
parameter for the light-rays). However, in the present context
the use of the proper time is not reasonable. Indeed, let us
consider the analogues of the first integrals
(\ref{eq:two})--(\ref{eq:three})--(\ref{eq:four}) with the proper
time $s$ in lieu of the affine parameter $p$; we have:

\begin{equation} \label{eq:sixteen}
\frac{r}{r-\alpha} \, \left(\frac{\textrm{d}r}{\textrm{d}s}
\right)^{2} +r^{2}  \left( \frac{\textrm{d}\varphi}{\textrm{d}s}
\right)^{2} - \frac{r-\alpha}{r} \, \left(
\frac{\textrm{d}t}{\textrm{d}s} \right)^{2} = -1 \quad;
\end{equation}

\begin{equation} \label{eq:seventeen}
r^{2} \, \frac{\textrm{d}\varphi}{\textrm{d}s} = L \quad;
\end{equation}

\begin{equation} \label{eq:eighteen}
\frac{r-\alpha}{r} \, \frac{\textrm{d}t}{\textrm{d}s}= E \quad ,
\end{equation}

where $L$ and $E$ are two constants. Formal deductions from
(\ref{eq:sixteen})--(\ref{eq:seventeen})--(\ref{eq:eighteen})
yield:

\begin{equation} \label{eq:nineteen}
\left(\frac{\textrm{d}r}{\textrm{d}s} \right)^{2} = E^{2}-
\frac{r-\alpha}{r^{3}} \, L^{2}- \frac{r-\alpha}{r}\quad ;
\end{equation}

\begin{equation} \label{eq:twenty}
\frac{\textrm{d}^{2}r}{\textrm{d}s^{2}} = \frac{L^{2}}{2} \,
\left( \frac{2}{r^{3}} - \frac{3\alpha}{r^{4}}\right) -
\frac{\alpha}{2r^{2}} \quad ;
\end{equation}

from which, when $r=\alpha$:

\begin{equation} \label{eq:twentyone}
\left(\frac{\textrm{d}r}{\textrm{d}s} \right)^{2}_{r=\alpha} =
E^{2} \quad ;
\end{equation}

\begin{equation} \label{eq:twentytwo}
\left(\frac{\textrm{d}^{2}r}{\textrm{d}s^{2}} \right)_{r=\alpha} =
\frac{1}{2} \, \left( \frac{L^{2}}{\alpha^{3}} -
\frac{1}{\alpha}\right) \quad ;
\end{equation}

However, eqs. (\ref{eq:twentyone}) and (\ref{eq:twentytwo}) are
actually meaningless: when $r=\alpha$, $\textrm{d}s^{2}=\infty$.
(In his memoir \cite{2} Hilbert did not use $s$ as an evolution
parameter!) -- Obviously, computations with an affine parameter
$p$ as evolution parameter would give finite, \emph{but}
$p$-\emph{dependent} values for $(\textrm{d}r / \textrm{d}p)^{2}_{r=\alpha}$ and $(\textrm{d}^{2}r / \textrm{d}p^{2})
_{r=\alpha}$.

\par All the geodesic lines of the light-rays satisfy the equation
$\textrm{d}s^{2}=0$, from which it is very natural to infer that

\begin{equation} \label{eq:twentythree}
\left(\frac{\textrm{d}r}{\textrm{d}t} \right)^{2} +
r\,(r-\alpha)\, \left(\frac{\textrm{d}\varphi}{\textrm{d}t}
\right)^{2}= \left(\frac{r-\alpha}{r}\right)^{2}\quad ,
\end{equation}

\begin{equation} \label{eq:twentyfour}
\left(\frac{\textrm{d}r}{\textrm{d}t} \right)^{2}_{r=\alpha} = 0
\quad ;
\end{equation}

\begin{equation} \label{eq:twentyfive}
\left(\frac{\textrm{d}^{2}r}{\textrm{d}t^{2}} \right)_{r=\alpha} =
0 \quad .
\end{equation}

But also for the test-particles the ``Systemzeit'' $t$ is the
\emph{unique} evolution parameter which gives always \emph{real
physical} results.

\par A last remark. In the Schwarzschildian \emph{original} form
of $\textrm{d}s^{2}$ ($(r^{3}+\alpha^{3})^{1/3}$ in lieu of $r$ in
eq. (\ref{eq:one})), or in Brillouin's form ($r+\alpha$ in lieu of
$r$ in eq. (\ref{eq:one})), the spatial region $0 \leq r <
\alpha$ is \emph{absent}. The manifold is \emph{maximally
extended}. Of course, \emph{all} the physical results which can be
derived from eq. (\ref{eq:one}) can be obtained also with
Schwarzschild's and Brillouin's metric forms. In particular,
\emph{the gravitational repulsion is a phenomenon of invariant
character}, \emph{i.e.} independent of the space-time reference
frame.

 \vskip1.20cm \noindent \textbf{4.} -- ``It si widely believed that the gravitational
  field of any electrically neutral collapsing body will eventually approach $[$by virtue
 of an assumed rotation$]$ the Kerr form.'' (Weinberg \cite{5}).

 \par Now, we have proved that a test-particle, or a light-ray,
 moving through the Kerr manifold along a radial geodesic in the
 negative direction of the radial coordinate arrive at the
 ``stationary-limit''surface with a zero three-velocity and a
 positive, or zero, three-acceleration: a clear instance of a
 Hilbert's repulsive effect, whose action we have computed for a
 generic value of the $\vartheta$-coordinate \cite{6}.

\vskip1.20cm \noindent \textbf{5.} -- In a recent paper \cite{7}
we have proved that if one takes into account the Hilbertian
gravitational repulsion, even a ``dust'' sphere with an internal
zero pressure collapses to a body of a \emph{finite} volume.
Precisely, the collapse ends when the mass $m$ of the sphere
has filled up the spatial region $0\leq r \leq \alpha (\equiv
2m)$. It follows that the gravitational field for $r>\alpha$ of
this (relatively small) spherical body coincides with the field of
a point-mass $m$. Both these objects have the physical property
that -- by virtue of the Hilbertian repulsion -- \emph{they are
incapable of ``swallowing'' anything}, light or material
corpuscles. (\emph{N.B.} -- \emph{For the adjustment of the internal to
the external solution we assume for simplicity that the radius of
the ``full globe'' is equal to} $2m+\varepsilon$\emph{, with an arbitrary small}
 $\varepsilon>0$).

\par Of course, the Euclidean formula for a spherical volume
$V=(4/3) \pi r^{3}$ does not hold for the volume $U$ of the ``full globe'';
 the difference between $V$ and $U$ becomes larger and larger with
 the increase of the mass $m$.

 \par It is instructive to give a generalization for $B\neq 0$ of
 the equations $(\textrm{d}r / \textrm{d}t)_{r=\alpha}=0=(\textrm{d}^{2}r / \textrm{d}t^{2})_{r=\alpha}$,
 which we have previously recalled for the radial
 geodesics $(B=0)$.

 \par We get from eqs. (\ref{eq:five}) and (\ref{eq:three}) (with
 $C=1$):

\begin{equation} \label{eq:twentysix}
\left( \frac{\textrm{d}r}{\textrm{d}t}\right)^{2} = B^{2} \left(
\frac{r-\alpha}{r}\right)^{2} \, \left( \frac{1+A}{B^{2}} -
\frac{A\alpha}{B^{2}} \, \frac{1}{r} -  \frac{1}{r^{2}} +
\frac{\alpha}{r^{3}} \right) \quad ,
\end{equation}

from which, when $B=0$:

\begin{equation} \label{eq:twentysixprime}
\left( \frac{\textrm{d}r}{\textrm{d}t}\right)^{2}_{B=0} =
\left(\frac{r-\alpha}{r}\right)^{2} \, \left(1+A \,
\frac{r-\alpha}{r} \right) \tag{26$'$} \quad .
\end{equation}

For $r=\alpha$ eqs. (\ref{eq:twentysix}) and (\ref{eq:twentysixprime}) give:

\begin{equation} \label{eq:twentyseven}
\left( \frac{\textrm{d}r}{\textrm{d}t}\right)_{r=\alpha} = \left(
\frac{\textrm{d}^{2}r}{\textrm{d}t^{2}}\right)_{r=\alpha} = \, 0 \quad .
\end{equation}

And from eq. (\ref{eq:twentysix}):

\begin{equation} \label{eq:twentyeight}
\pm \, B \, \textrm{d}t = \frac{r}{r-\alpha} \,
\frac{\textrm{d}r}{\{[(1+A)/B^{2}]-(A\alpha/B^{2}r)-1/r^{2}+\alpha/r^{3}\}^{1/2}}
\quad ,
\end{equation}

which tells us that test-particles and light-rays reach
$r=\alpha$ after an infinitely long time: a result that is
universally known, but whose real meaning is usually neglected,
because of the preference given to the proper time $s$. However,
this preference is here mathematically \emph{baseless}, as we have
proved in sect. \textbf{3bis}.

\par Remark that if you put in eq. (\ref{eq:twentyeight})
$r=(9/8)\alpha$, \emph{e.g.}, instead of $r=\alpha$, you get a
reasonable time interval $\Delta t$.

\vskip1.20cm \noindent \textbf{5bis.} -- Under given conditions, the Hilbertian gravitational repulsion
in Schwarzschild's and Kerr's manifolds manifests itself in \emph{all} the geodesic paths, in particular
in the circular orbits and in the trajectories which arrive on the spatial surface $r=2m$. Due to an undue
 preference for the proper time with respect to the
 coordinate-time (\emph{i.e.}, the \emph{time of the system}), only
 the circular orbits are commonly believed to be subjected to
 Hilbert's repulsive action. This limitation is quite illogical:
 indeed, it is clear that the gravitational repulsion does not
 suspend its action for the non-circular geodesics, \emph{in
 primis} for the geodesics which ``strike'' the surface $r=2m$.

\vskip1.20cm \noindent \textbf{6.} -- Back to the paper by Broderick \emph{et al.} \cite{1}. The spherically-symmetric
 point-mass $m$ (sect.\textbf{3}) and the ``full globe'' $0\leq r
 \leq2m$  (sect.\textbf{5}) create an identical gravitational
 field in the external region $r>2m$ -- and both these gravitating
 objects give a gravitational repulsion under the illustrated
 conditions. For a diagram of some geodesics of light-rays, see
 Fig.1, that represents a correct mathematical counterpart (in a
 generic plane) of Fig.1-\cite{1}, which is reproduced with its
 legend in the following Fig.2. We emphasize that the authors
 restrict the gravitational repulsion to the unique circular
 orbit of the light-rays.

\begin{figure}[!ht]
\begin{center}
\includegraphics[width=0.6\textwidth]{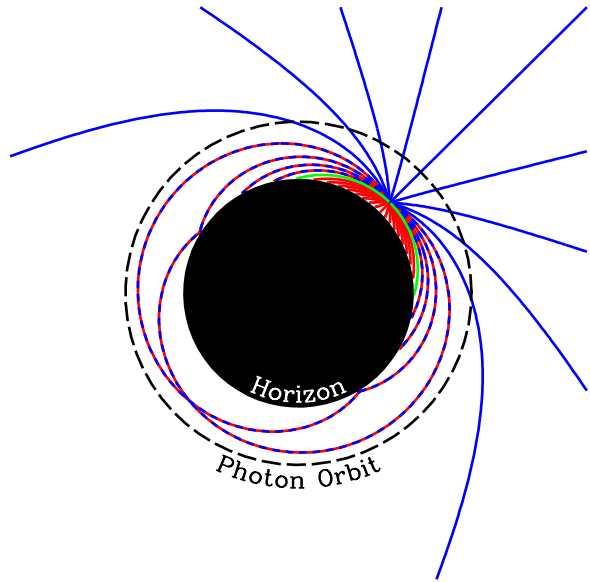}
\caption{\small(From Broderick \emph{et al.} \cite{1}).
Rays launched isotropically (every $10^\circ$) in the locally
flat, stationary frame are lensed in a Schwarzschild spacetime.
Those rays that are initially moving inwards, tangentially and
outwards are shown in red, green and blue, respectively.
Additionally, those that are launched initially moving outwards
and are subsequently captured are red-blue dashed.  For reference
the horizon and photon orbit are shown.  Generically, the fraction
of rays that escape to infinity decreases as the emission point is
moved towards the black hole, dropping below 50\% at the photon
orbit and dropping all the way to 0\% at the horizon.  As a
consequence of this strong lensing, emitting objects that are
contained within the photon orbit approximate the canonical
pin-hole cavity example of a blackbody, becoming a perfect
blackbody in the limit that the surface redshift goes to
$\infty$.}
\end{center}
\end{figure}
\normalsize

By virtue of the Hilbertian repulsive effect, the materials of the accretion flow arrive at $r=2m$
 with a zero velocity and a zero acceleration. Accordingly, the ``hollow
 globe'' $r<2m$ around the point-mass $m$ cannot ``swallow''
 anything. The accretion materials (matter fragments and light)
 perform a very soft landing on the ideal surface $\Im$ of the ``hollow
 globe'',  or on the physical surface $\scriptstyle \sum$ of the ``full globe''
 $0\leq r \leq 2m$. It seems to us that these facts explain the
 observational data of \cite{1}; in particular, it is obvious that
 the surfaces $\Im$ and $\scriptstyle \sum$ have a \emph{low}
 luminosity.

\par Quite generally, the effects of the Hilbertian gravitational repulsion suffice to cancel
 all the widespread (and less widespread) convictions about the exotic
 properties of the ideal surface $r=2m$.

\newpage
\begin{center}
\noindent \small \emph{\textbf{APPENDIX A}}
\end{center} \normalsize

\vskip0.40cm \noindent In recent years, some authors have emphasized that there are observational proofs
 of the existence of \emph{intrinsic magnetic moments} in
 BH-candidates -- both of stellar masses and AGN-masses --, for
 instance in SgrA*. Now, the existence of a magnetic moment (of an
 appreciable magnitude) \emph{forbids} the existence of the event
 horizon around a point-mass. The above authors have developed a
 sophisticated model of the collapse of massive and supermassive magnetic bodies
 with the purpose to offer an alternative explanation of the data
 by Broderick \emph{et al.}; in particular, they give a special
 prominence to the action of a magnetic propeller driven outflow
 for explaining the low bolometric luminosity of SgrA*.

 \par We think, however, that the reliability of this model (a
 heterogeneous offspring of GR and QED) is not evident.

\vskip2.00cm
\begin{center}
\noindent \small \emph{\textbf{APPENDIX B}}
\end{center} \normalsize

\vskip0.40cm \noindent It is useful to compare Droste's \cite{4} and Hilbert's \cite{2}
treatments about attraction and repulsion in the radial geodesic motions through Schwarzschild manifold.

\par Droste starts from this definition of radial geodesic
velocity:

\begin{equation} \label{eq:B1}
\frac{\textrm{d}\delta}{\textrm{d}t} :=
\frac{\textrm{d}r}{\textrm{d}t} \, \left(1-\frac{\alpha}{r}
\right)^{-1/2} \tag{B1} \quad , \quad (\alpha\equiv 2m) \quad ,
\end{equation}

which is suggested by the metric radial interval $\textrm{d}\delta
:= \textrm{d}r (1-\alpha r^{-1})^{-1/2}$. Eq. (\ref{eq:two}) of sect.\textbf{3} gives the first integral

\begin{equation} \label{eq:B2}
\left( \frac{\textrm{d}r}{\textrm{d}t} \right)^{2} =
\left(1-\frac{\alpha}{r}\right)^{2} \, \left[1+A \, \Big(1-\frac{\alpha}{r}\Big) \right]  \tag{B2} \quad
,
\end{equation}

from which

\begin{equation} \label{eq:B3}
\left( \frac{\textrm{d}\delta}{\textrm{d}t} \right)^{2} =
\left(1-\frac{\alpha}{r}\right) \, \left[1+A \,
\Big(1-\frac{\alpha}{r}\Big) \right]  \tag{B3} \quad .
\end{equation}

As a consequence of

\begin{equation} \label{eq:B4}
\frac{\textrm{d}^{2}r}{\textrm{d}t^{2}} =
\frac{3\alpha}{2r\,(r-\alpha)}  \,  \left(
\frac{\textrm{d}r}{\textrm{d}t} \right)^{2} - \frac{\alpha \,
(r-\alpha)}{2r^{3}}
 \tag{B4} \quad ,
\end{equation}

which can be easily derived from eqs. (\ref{eq:four}) and
(\ref{eq:six}) of sect.\textbf{3}, Droste arrives at the following
expression for his acceleration $\textrm{d}^{2}\delta /
\textrm{d}t^{2}$:

\begin{equation} \label{eq:B5}
\frac{\textrm{d}^{2}\delta}{\textrm{d}t^{2}} =  -
\frac{\alpha}{2r^{2}} \, \left[ \Big(1-\frac{\alpha}{r}\Big)^{1/2}
- \frac{2\,(\textrm{d}\delta / \textrm{d}t)^{2}}{\Big(1-\frac{\alpha}{r}\Big)^{1/2}} \right]
 \tag{B5} \quad .
\end{equation}

Let us call $r_{*}$ the value of $r$ for which $\textrm{d}^{2}\delta / \textrm{d}t^{2}=0$
(attraction counterbalances repulsion); we have:

\begin{equation} \label{eq:B6}
\left|\frac{\textrm{d}\delta}{\textrm{d}t}\right|_{r=r_{*}} =
\frac{1}{\sqrt{2}} \, \left(1-\frac{\alpha}{r_{*}}\right)^{1/2}
\tag{B6} \quad ,
\end{equation}

from which:

\begin{equation} \label{eq:B7}
\left|\frac{\textrm{d}r}{\textrm{d}t}\right|_{r=r_{*}} =
\frac{1}{\sqrt{2}} \, \left(1-\frac{\alpha}{r_{*}}\right)
\tag{B7} \quad ;
\end{equation}

but if $r^{*}$ is the value of $r$ for which $\textrm{d}^{2}r
/ \textrm{d}t^{2}=0$, we have with Hilbert \cite{2}:

\begin{equation} \label{eq:B8}
\left|\frac{\textrm{d}r}{\textrm{d}t}\right|_{r=r^{*}} =
\frac{1}{\sqrt{3}} \, \left(1-\frac{\alpha}{r^{*}}\right)
 \tag{B8} \quad ;
\end{equation}

we see that there is a non-negligible difference between Droste's B.(6)--(B.7) and Hilbert's (B.8) maximal velocities.

\par We think that \emph{Hilbert's treatment must be preferred},
for the following reason.

\par The metric distance $\delta$ of the point $(r, \vartheta,
\varphi)$ from $r=\alpha$ is:

\begin{equation} \label{eq:B9}
\delta = \int^{r}_{\alpha} \left(1-\frac{\alpha}{r'}\right)^{-1/2}
\textrm{d}r' = r \, \left(1-\frac{\alpha}{r}\right)^{1/2} + \alpha
\ln \left| \Big(\frac{r}{\alpha}-1 \Big)^{1/2} + \Big(\frac{r}{\alpha}\Big)^{1/2} \right|
 \tag{B9} \quad ;
\end{equation}

by virtue of the arbitrariness in the choice of the coordinates,
in particular of the radial coordinate, if we put
$f(r):=\delta(r)+\alpha$, and perform in eq. (\ref{eq:one}) of
sect.\textbf{3} the substitution $r\rightarrow f(r)$, we get a $\textrm{d}s^{2}$
which is physically equivalent to the one of eq. (\ref{eq:one}). From the mathematical standpoint,
 this new expression of the $\textrm{d}s^{2}$ is
 \emph{diffeomorphic} to that of eq. (\ref{eq:one})
 \emph{for} $r>\alpha$. Now, let us choose a definition \emph{\`a la}
 Hilbert for the radial velocity: we have $\textrm{d}f(r) / \textrm{d}t= \textrm{d}\delta(r) / \textrm{d}t$,
 \emph{i.e.} Droste's definition (B.1).

 \par However, all the consequences coincide now with those of Hilbert's treatment; in particular,
 $\textrm{d}^{2}\delta(r) / \textrm{d}t^{2}$ is now equal to zero if and only if $\textrm{d}^{2}r / \textrm{d}t^{2}=0$.
 We see that a $\textrm{d}s^{2}$ expressed \emph{in toto} with
 the metric coordinate $f(r)=\delta(r)+\alpha$ gives the
 same results of the $\textrm{d}s^{2}$  of eq. (\ref{eq:one}) in Hilbert's procedure.

\end{document}